\documentclass[a4paper,12pt]{article}
\usepackage{epsfig}
\usepackage{amssymb}
\usepackage{graphicx}
\usepackage{amsmath}

\begin{document}

%\tolerance=100000

\thispagestyle{empty}

\setcounter{page}{0}

\vspace*{\fill}

\begin{center}

{\Large \bf First scientific results from the Estonian Grid}\\[2.cm]

{{\large\bf Andi Hektor}$^1$,
{\large\bf Lauri Anton}$^2$,
{\large\bf Mario Kadastik}$^1$,
{\large\bf Konstantin Skaburskas}$^3$ {\large and}
{\large\bf Hardi Teder}$^2$}\\[7mm]

{\it $^1$ National Institute of Chemical Physics and Biophysics,\\
R\"avala pst 10, 10143 Tallinn, Estonia}\\[3mm]
{\it $^2$ Estonian Educational and Research Network,\\
Raekoja plats 14, 51004 Tartu, Estonia}\\[3mm]
{\it $^3$ Institute of Technology, University of Tartu,\\
Vanemuise 21, 51014 Tartu, Estonia}\\[3mm]

\end{center}

\vspace*{\fill}

\begin{abstract}
{\small\noindent We present first scientific results, technical details and recent developments in the Estonian Grid. Ideas and concepts behind Grid technology are described. We mention some most crucial parts of a Grid system, as well as some unique possibilities in the Estonian situation. Scientific applications currently running on Estonian Grid are listed. We discuss the first scientific computations and results the Estonian Grid. The computations show that the middleware is well chosen and the Estonian Grid has remarkable stability and scalability. The authors present the collected results and experiences of the development of the Estonian Grid and add some ideas of the near future of the Estonian Grid.}
\end{abstract}

\vspace*{\fill}

\newpage
\setcounter{page}{1}

\section{Introduction}

Since the beginning of the computer age, supercomputers have been at the forefront of scientific computation due to high resource and computational needs. In the past years this approach has changed. Since building a stable supercomputer is a very expensive task, more and more is invested in search for cheaper and more flexible alternatives \cite{InfFut:1994}. On the other hand, many international scientific experiments will need a huge computational power in near future and it will be impossible in practice to cover these needs using any one supercomputer. Additionally, nowadays the communities and experimental facilities of international scientific experiments are distributed geographically. The Grid technology promises to give solution for the both sides \cite{Foster:2001}. It is an ideology of using low cost personal computer clusters as the building blocks and the Internet to connect the blocks. By interconnecting them through so called Grid middleware we can put together a  "virtual" supercomputer \cite{Foster:2003}.

In modern terms, the Grid is a standardised layer of software between the cluster query systems, storage elements, users and other Grid resources. Grid middleware is the component that will bind all kinds of resources (different hardware architecture, different operating systems, etc) into a uniform system with standardised tools. Its power lies in parallelisation and massive execution of computational tasks, so called Grid jobs. It enables scientists to create a computational job, split it into as many independent sub-jobs and then send them to the Grid. Finally,
after the jobs are calculated the results are downloaded and analyzed. That kind of parallelisable computation is most suited for massive scientific calculations like detector simulations and data analysis in high energy and radiation physics, genetics and bioinformatics, climate modelling and weather forecast, but it is also suited for many commercial purposes, for example rendering of animations in movie industry, simulate electronic systems, etc. 
A deeper overview of Grid ideology and technology is given in the books by Foster et al \cite{Foster:2003} and Berman et al \cite{Berman:2003}.

There are many projects under development leading to a diversity of technological approaches, for example UNICORE, Globus, Legion, Gridbus, etc. Same of them support only a minimal set of functionality (e.g. Globus), others try to develop a maximum set of tools and functions (e.g. UNICORE) \cite{Asadzadeh:2004}.

Some international scientific collaborations will very soon need Grid technology: LHC (Large Hadron Collider) at CERN (European Organization for Nuclear Research) \cite{CERNexp}, the Planck Mission for analysis of cosmological background radiation \cite{PlancMis}, analysis of the human genome at the Human Genome Project \cite{HumanGenomeProject} etc. CERN is a leading developer and user of Grid technology due to the schedule of the LHC experiment (starting in 2007) \cite{CERNandGrid}.

It is important to mention that Grid system does not cover the functionality of supercomputer one-by-one. On one hand, Grid has many additional possibilities for large-scale collaboration, but on the other hand it does not have several functionalities of a typical supercomputer. The strongest restriction is the inability to use shared or distributed memory parallel computing in geographically distributed Grid environment. This restriction comes from the latency of Internet connection. Finally, the speed of light sets a natural limit for shared memory on the geographical scale.

In the scientific part of the paper we focus on two tasks. First we propose a new method for modelling same spectra in the mobile and small gamma spectrometer: we combine the Monte Carlo (MC) method in radiation physics and distributed calculations on Grid. Due to experimental complications and expenses the numerical modelling is a suitable method to model gamma spectrometers. Unfortunately it needs a lot of computational resources. In the Grid we can divide our MC simulation into hundreds of independent sub-simulations and send them to the Grid. It is a practical way for computations, but it a good possibility to test a Grid system as well. We used the statistics of the sub-jobs to estimate the stability and reliability of the system.

The using of Grid was very promising, the Grid middleware used was sufficiently stable for this type of computations. The results of the MC modelling are realistic and give some needed hints for the experimental set up in the future.

\section{Architecture of Grid}

To ensure unified standards the Global Grid Forum (GGF) sets the specifications and agreements for Grid systems: OGSA (Open Grid Services Architecture), RSL (Resource Specification Language), etc \cite{GGF}. As mentioned, not all Grid projects follow the specifications precisely and there are some freedom and uncovered areas in the specifications and agreements.

\begin{figure}
   \centering
   \includegraphics{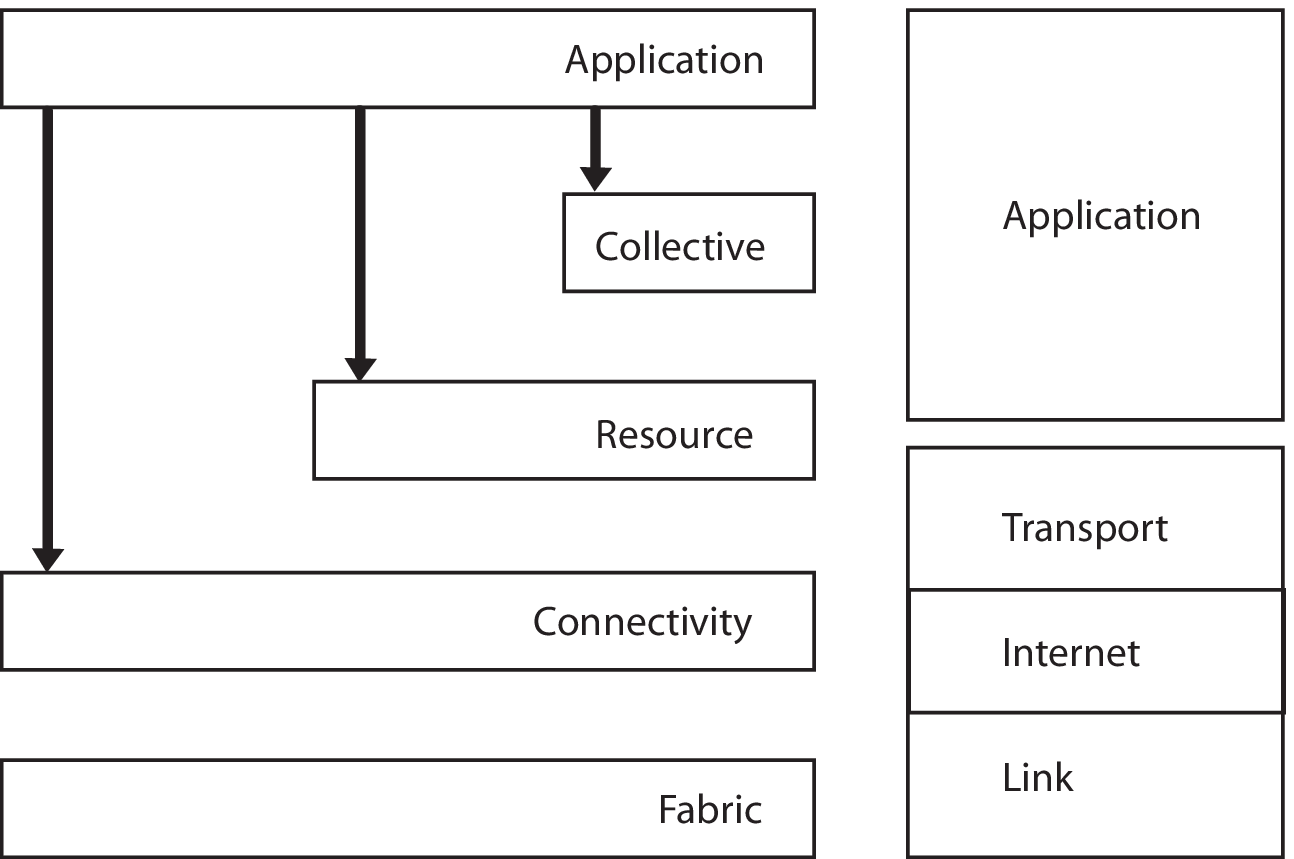}
   \caption{The logical layers of the Grid connected to the layers of the Internet.}
   \label{fig:GridLayers}
\end{figure}

Figure \ref{fig:GridLayers} presents the layered architecture of a typical Grid system \cite{Foster:2001}. For a fully functional Grid one needs four components (named by the GGF \cite{GGF} and NorduGrid project \cite{NorduGrid}):
\begin{itemize}
    \item Cluster Element (CE) -- an
    actual node/farm/cluster/supercomputer on what Grid jobs are
    executed. Typically in a Grid environment this is a Linux farm
    with some job scheduling system like PBS. The frontend of the
    farm is connected to Grid where it accepts Grid jobs and
    submits them as local jobs. Once the jobs have finished it will
    return the results as specified.
    \item Storage Element (SE) -- a node that has some data storage
    resources attached to it that is available for Grid usage. As the Grid is a
    decentralised system, there is need to store input and
    result files in a common place for easier management. Special
    storage elements were designed for this purpose that allow
    storage and retrieval of files through Grid.
    \item QDRL (Quasi Dynamic Resource Locator), also known as
    GIIS (Grid Index Information Service) -- is responsible for
    information propagation of Grid resources through the entire
    network. CE and SE register their resource information to
    the local GRIS (Grid Resource Information System) which is typically
    located in the CE or SE itself. GRIS registers its address in
    the nearest QDRL which then propagates the GRIS location up
    its chain.
    \item UI (User Interface) -- a command line or GUI interface
    allowing user to submit new Grid jobs, monitor existing ones,
    retrieve results, kill running jobs, etc.
    \item Special Component or Device -- a special facility connected directly to the Grid, for example a particle detector, an environmental sensor, etc.
\end{itemize}

A typical Grid middleware toolkit, the Globus Toolkit, is produced by the Globus project \cite{GlobusToolkit}. Globus Toolkit (GT) is in itself not a fully functional Grid middleware, but a Grid middleware toolkit/API. It is popular: many middleware packages have been built upon GT and rely upon its functions for the basic functionality like GSI (Grid Security Infrastructure), MDS (QDRL is based on MDS, which is a modified LDAP system), Globus-IO (for file transfers) etc. Globus project itself is a collaboration connecting many scientific institutions and sponsored by many organisations like NASA, IBM, Cisco, Microsoft etc \cite{GlobusToolkit}.

\subsection{Security of Grid}

Also a very important aspect of Grid is the ability to submit and download jobs in secure way from anywhere in the world. Most Grids under development around the world use Public Key Infrastructure (PKI) as a method for authentication the actors of the Grid \cite{GridAndPKI}. Following PKI every user and resource element in the Grid has to have a certificate signed by Certification Authority (CA). The using of PKI gives a unique possibility in Estonia. It is possible to use the local electronic ID-card (SmartCard) infrastructure in Estonia based on PKI for the potential Grid users \cite{EstIDcard}. It means that an Estonian inhabitant with valid electronic ID-card can use that for the Grid. Thus, there is a possibility to save resources using this ready PKI structure. There are only some countries in Europe where the electronic PKI-based ID-card infrastructure are available: Belgium, Estonia, Finland and Sweden \cite{Martens}.

\section{Development of the Estonian Grid}

It is impossible to mark the exact moment of the birth of the Estonian Grid (EG). Therefore, the authors give the list of the memorable dates:
\begin{itemize}
    \item {\bf Jan-Dec 2003}: Some coordinative meetings at the Estonian Educational and Research Network (EENet), National Institute of Chemical Physics and Biophysics (NICPB), University of Tartu (UT) and Tallinn University of Technology.
    \item {\bf Jan 2004}: The first components of the EG: EG CA, $\beta$-level GIIS, the first computer in the EG established at NICPB. Technical meeting of NorduGrid at the NICPB in Tallinn, the first public seminar of the EG.
    \item {\bf Feb 2004}: Establishing of the Centre of High Energy Physics and Computational Science at NICPB for the purpose to support the development of the local Grid applications for high energy and radiation physics and material science.
    \item {\bf Feb-March 2004}: First multiprocessor clusters in the EG at the EENet and UT. The public Web page of the EG \cite{EGWeb}
    \item {\bf April 2004}: The first draft of Certification Policy and Certification Practice Statement document for the EG CA. The first regular seminars of Grid technology at the NICPB and UT
    \item {\bf May 2004}: The collaboration protocol between CERN, LHC Computing Grid Project (LCG) and the Republic of Estonia was signed. The first scientific software ported to the EG from the authors \cite{Kadastik:2004}. The establishing of the Grid Laboratory at the Institute of Technology of the UT.
    \item {\bf June 2004}: Establishing of EG technical coordination and supporting group \cite{EGTechSup}. The first massive tests on the EG.
    \item {\bf July 2004}: The first massive scientific calculations on the EG.
\end{itemize}

The authors have participated and made presentations in the various international meetings and conferences: the NorduGrid meetings in Lund, Tallinn and Helsinki and the CACG Florence Meeting in Florence in 2004.

\section{Technical details and problems}

\subsection{Choice of middleware and compatibility}

The first problem of a potential developer of a Grid system is to find functional middleware for that. As mentioned before there are many different possibilities, but most middleware packages are in testing state with poor support for the users. We followed the tendency of CERN \cite{CERNandGrid} and our neighboring countries \cite{NorduGrid} and decided to use the middleware based on Globus Toolkit. One of the well-supported and functional middlewares based on the Globus Toolkit, the ARC (Advance Resource Connector) middleware, is developed by the NorduGrid project \cite{NorduGrid}. That is the first and foremost reason why this middleware is used for the EG.

The strongest alternatives of the ARC software were the EDG package (based on Globus Toolkit) \cite{EuropeanDataGrid} and UNICORE \cite{UNICORE}. The problem of EDG is insufficient user support. The problem of UNICORE is that it is unsupported by CERN and some other international projects.

On level of OGSA all the middleware packages are compatible. However, there are many non-compatible sub- or additional components in different Grid systems, e.g. the schema of information system, management of storage elements, etc. Hopefully, these non-compatibilities will soon be resolved.

\subsection{Grid PKI}

Traditionally each country has at least one CA for Grid. (There are some exceptions, e.g. NorduGrid project \cite{NorduGrid}). The CA of the EG follows the X.509 rules \cite{X.509}. The Certification Policy and Certification Practice Statement (CP/CPS) document for the CA is composed. It is presented online on the web page of the EG \cite{EGCA}. The CP/CPS document follows all the EUGridPMA rules, it is synchronized with the EUGridPMA organisation \cite{EUGridPMA} and the local CA is trusted by the EUGridPMA members. In the near future, we will support the Estonian electronic PKI-based ID-card infrastructure \cite{EstIDcard} for the EG.

\subsection{Information System and Web-based monitoring}

The information system of the EG is based on the NorduGrid LDAP schema \cite{InfoSystemOfNorduGrid}. It is decentralised and follows a structure that is similar to the Internet DNS. There are typically three levels of information servers or QDRLs: the top level QDRLs (so called $\alpha$-level), the second (or country) level QDRLs
(so called $\beta$-level) and the third or unit level QDRLs (so called $\gamma$-level). The tree of QDRL can be continued with an additional lower level QDRLs if needed.

The information coming directly from LDAP is not really human readable. Therefore, the Web-based monitoring interface is developed by the NorduGrid project. There are many online monitors available, for example the NorduGrid Monitor \cite{NGMonitor}, EG Monitor \cite{EGMonitor}, etc.

\subsection{Present state of the EG}

Figure 2 gives a technical view of the development of the EG. It shows some important parameters of the EG: the total number of CPUs and total RAM connected to the EG. Currently the EG has 7 CEs with 62 CPU in total. Most CPUs are from Intel (between 2.4 and 3.06 GHz) and the average RAM is between 512 MB and 1 GB. There are two test SEs available with 160 GB storage in total in the EG. The exact numbers and current situation of the resources are presented by the EG Monitor \cite{EGMonitor}. A typical link between the PCs of a CE is 1 Gbps. The fast Myrinet (2 Gbps) link is used at the NICPB cluster and the fast Scali link is used at the EENet cluster. The links between the CEs are limited by the typical Internet connection in Estonia (below 1 Gbps).

\subsection{General problems}

Some problems are recognized during the building up of the EG. There are two big classes of the problem: organisational and technical problems. The organisational problems are caused by the fact that the Grid is distributed system. There is a similarity between the beginning of the Internet and Grid \cite{Foster:2001}.
Special decisions are needed for resource sharing and local and general responsibilities. It can be solved by the special contracts between resource owners and/or personal agreements between technical persons. At the moment all the problems have been solved on the level of technical and personal agreements. However, in near future the EG will need some official agreements between resource owners.

The technical problems are connected to the fact that Grid middleware is in very rapid developing process. Generally, our choice, the NorduGrid ARC package has shown surprisingly good stability. There are much more problems connected to management of the hard- and software of the clusters than the problems connected to the middleware.

\section{Scientific applications ported to the EG}

\subsection{Modelling and analysis software for the international CMS experiment}

The CMS (Compact Muon Solenoid) experiment \cite{CMS} involves one of the biggest high energy particle detector at the LHC  (Large Hadron Collider) at CERN scheduled to begin operation in the year 2007. The LHC will collide 7 TeV proton beams head on. It can also collide beams of heavy ions such as lead with total collision energy in excess of 1250 TeV. The main objective of the CMS and LHC is explain the origin of particle mass, flavour and possible unification of fundamental interactions at high energy scales. In the Standard Model of particle interactions all the charged fermions acquire masses due to the spontaneous breaking of gauge symmetry via the Higgs mechanism, predicting the existence of a physical neutral scalar particle, the Higgs boson.

The CMS collaboration involves about 1990 scientists coming from 150 institutions distributed in 31 nations. The NICPB have been in collaboration since 1996. One of our tasks is to study and port of CMS software to simulate, digitize and analyse event creation in the CMS detector. The software consists of three parts:
\begin{itemize}
\item CMKIN \cite{CMSOO} -- a Monte Carlo (MC) physics generation application written to simplify the use of different MC generators like PYTHIA \cite{PYTHIA}, HERWIG, TAUOLE, TopRex etc. It is used to generate ideal proton-proton collisions and the produced particles and its output is taken as input to OSCAR.
\item OSCAR \cite{CMSOO} -- simulates particle passage through the CMS detector and simulates hits in different parts of detectors. OSCAR uses the Geant4 software package that is described in the next subsection.
\item ORCA \cite{CMSOO} -- the actual tool that will be used also when the real detector goes online. It is currently used for data reconstruction from simulated runs and also for later analysis.
\end{itemize}

One event in the above software means one proton-proton collision within the CMS detector. The collisions will happen approximately $10^8$ times per second when the LHC will go online. It means that the data production of the detectors of the LHC will be huge, about 10~000~TB data per year. The analysis of this data needs computational power that is equal to about 100~000 fast modern PCs. Fortunately the events can be looked at as separate entities as they do not depend on other events and the Grid can be the solution for the data analysis.

The current tests that have been performed on the EG have been the creation of data sets from CMKIN particle generation to ORCA reconstruction and also some preliminary analysis. The code has worked remarkably well and we have managed to produce a lot of events for later more detailed analysis.

\subsection{Parallel solver for linear systems of equations}

DOUG (Domain decomposition On Unstructured Grids) is a black box parallel iterative solver for finite element systems arising from elliptic partial differential equations \cite{DOUG}. Used in conjunction with a finite element discretisation code, DOUG will solve the resulting linear systems using an iterative method, and provides a range of powerful domain decomposition preconditioners.

The code is designed to run effectively in parallel on virtually any machine that supports MPI. The matrix-vector operations arising in the iterative method are parallelised using graph partitioning software and
additive Schwarz preconditioners can be automatically constructed by the DOUG using only minimal input. A full additive Schwarz preconditioner with automatically generated coarse grid is provided in 2D and 3D. The DOUG
makes no assumptions whatsoever about the finite element mesh that the problem arises from; it may be as unstructured as necessary and only the basic output from the mesh generator and the finite element discretisation are required as inputs to the DOUG.

Currently the DOUG is mainly used in solution of matrices having block structure which arise from discretisation of coupled systems of different differential equations (like the Navier Stokes flow equations) and in stability assessment problem of nuclear power station cooling systems. This research is done in collaboration with researchers from Bath University (Spence~A. and Graham~I.G.) and AEA Technology (Cliffe~K.A.) in UK.

The DOUG has a graphical user interface implemented as a Web-interface \cite{DOUGWeb, DOUGamst}. The Grid-awareness is added to the Web-interface for the DOUG and it is available for the EG users \cite{DOUGGrid}. During the development of Grid-enabled Web-interface the problem of action on the Grid by the interface on behalf of the user and necessity of managing users' credentials -- Grid-proxies -- had arisen. Those issues were successfully solved by using MyProxy (Online Credentials Repository \cite{MyProxy}) and appropriate developing and coding of the
interface. We have two MyProxy servers installed on EG \cite{EGMyProxies}.

\subsection{Radiation and particle physics}

Geant4 is a software toolkit for the MC simulation of the passage of particles through matter
\cite{Geant4}. Its application areas include high energy physics and nuclear experiments, medical, accelerator and space physics studies. Geant4 covers the energy scale from 250 eV ut to some TeV for most known particles and interaction processes.

The Geant4 is used like a external library for many software packages: the OSCAR software mentioned above, medical software for radiotherapy for cancer treatment, etc. The Geant4 software was developed by RD44 group \cite{RD44}, a world-wide collaboration of about 190 scientists participating in more than 10 experiments and 20 institutions in Europe, India, Japan, Canada and the United States.

There are three collaboration projects using Geant4 in environmental physics, medicine and particle/radiation detectors in Estonia. Therefore we are interested in supporting and using the Geant4 software at the EG. The first scientific calculations on the EG have been done using the Geant4 software.

\subsection{Coming scientific and non-scientific applications}

There is one group (Karelson~M {\it et al}) using and developing the UNICORE middleware for the OpenMolGRID (Open Computing Grid for Molecular Science and Engineering) project in Estonia \cite{OpenMolGRID}.

Additionally many other work groups in Estonian science and technology are interested in Grid technology and the EG: analysis of gene information, climate modelling, material science, etc. The interest is arising in the commercial sphere as well. Some companies need computational power in the different topics: material engineering, nuclear safety, computer animations, military applications, etc.

\section{First scientific challenge of the EG}

\subsection{MC simulations in radiation physics}

For the first massive test of the EG we made some intensive scientific computations using the Geant4 software package. In our study MC method is used to model the operation of a scintillation detector installed in a prototype radiation surveillance unit on the small unmanned airplane, the so called Ranger. Planned measurements by Ranger are complicated and dangerous for the humans (e.g., radioactive cloud, etc.), thus, the detection capability has to be estimated by calculative means. Most importantly the limits of the detector have to be estimated \cite{STUK:2003}.

In the study we analyse the simplest case: a isotropic radioactive point source on the ground and the detector directly above the source in the different heights. A very practical question for radiation surveillance is the difference of the spectra between different heights. We modelled the spectra at the heights 150~m and 100~m (Figure \ref{fig:stuk-system}).

We have to simulate billions of gamma events to get statistically good histograms. Therefore the calculations are very time consuming. The number of registered gamma quants N follows the approximate equation:
\begin{equation}
N \propto \frac{1}{r^2}\exp(-\mu r),  \label{v1}
\end{equation}
where $r$ is the distance between the detector and the point source, the attenuation coefficient $\mu$ of the environment between the source and detector. Thus, the statistical uncertainty of N is:
\begin{equation}
err_N = \sqrt{N} \propto \frac{1}{r}\exp(\frac{-\mu r}{2}).  \label{v2}
\end{equation}

Luckily an intrinsic property of radioactive emission events (and the MC calculations as well) are that all the events are independent and we can split the computations to smaller subtasks. So, this task is excellent for testing purposes of a Grid system. In the following studies we focus mostly on the testing of the EG. The experimental studies, exact details of the calculations and additional studies will be published separately \cite{STUKHektor:2004}.

We built up our model in Geant4 (release 5.2), compiled and made some test runs locally on a Linux PC (2.8~GHz Intel Pentium4, 1~GB RAM, RedHat 9.0). Typical compilation time of the source code of the model was around some minutes. If the distance between the point source and the detector is between 100~m and 150~m then the calculation time is between 0.05~ms and 0.2~ms per a source event.

The second step was to send the computations to the EG. There are two possibilities to send a Grid job to the Grid. First, we can send the source file of our code to the Grid and it compiles and runs on a Grid node (CE). It means the external libraries (e.g., Geant4) has to be installed to the CE before the sending of the job. Second, we can compile our code locally and then send the compiled binary to run on a CE of the Grid. The drawback of the last case is that the CE has to have the suitable operation system, correct version of glib, etc. In addition, if the binary file is big then a lot of time is spent on uploading the file. In the case of our Geant4 radionuclide detection simulation, the binary file is rather small (3.7~MB), it runs on the most CEs in the EG and therefore we prefer the last variation.

We simulated all the possible combinations: two different radionuclide ($^{137}$Cs and $^{60}$Co), two different detector positions (100~m and 150~m) and additionally we changed a parameter of the system, the radius of the air cylinder around the system (we used two different radiuses 40~m and 100~m). In sum, there are 8 different cases. For satisfactory statistics in each case 80 billion events were simulated. In total it means $8 \times 80 = 640$ billion events. We divided the total set of the simulations (640 billion) in sets of 1 billion events. The computation time of a set is reasonable (8-16 hours) and in compliance with the recommended maximum cycle of the random number generator of Geant4 \cite{Geant4,CLHEP}. All 640 sets were submitted as Grid jobs to the EG.

The Figure \ref{fig:stuk-Cs} presents the results of the simulations for $^{137}$Cs and Figure \ref{fig:stuk-Co} for $^{60}$Co. We present the spectra only for the radius of 40~m. There is only a very slight difference between the 40~m and 100~m radiuses. It is very close to the error limit and therefore we do not present the curves of the 100~m radius in the figures. We can see clear difference between two detector positions, 150~m and 100~m. To estimate the concrete detection limit we have to compare these curves with the local natural radioactivity background case-by-case. The line structure in the Compton continuum region needs additional studies.

\subsection{Reliability of the EG}

The speed of the CPUs used were between 2.4~GHz and 3.06~GHz (Intel Pentium 4) and the available RAM per PC was between 512 MB and 1~GB. The total time of the computations was 417 CPU days.

The results were very promising. No jobs failed due to the Grid middleware. In total there were 17 failed jobs (2.6\%) during the computations, probably due to random hardware errors. These jobs were resubmitted to the EG and they finished successfully. There were 16 post-processing errors (2.5\%) due to the instability of the hardware/software caused by the external factors: blackouts of electric grid, overheating, etc.

However we only tested some functionalities of the Grid. The stability of the data management and runtime environments need additional testing.

\section{Summary and some conclusions}

The first experiences and tests of the EG have been promising. The development of the EG has been impressive, especially if we compare the results and the spent resources. The installation of the middleware, management of middleware and management of the CA of the EG, certificates, QDRLs, EG Monitor and some CEs/SEs are done mostly as volunteer work.

The first scientific calculations on the EG show that it is a very useful tool for the computational scientists, especially in the field of computational particle and radiation physics. If the authors compare the earlier experiences of the computations on the PC clusters \cite{HektorAarhus, HektorUppsala} then the using of the Grid simplifies substantially the performing of the scientific computations.

During the first year of the EG the authors have experienced many technical and organizational problems and bottle-necks, presented here in short.
\begin{itemize}
    \item {\bf Instability of the hardware of CE.} It is a typical problem for the managers of PC farms and clusters and parallel computers. In our case it was mostly caused by electric blackouts or overheating. Using the Grid can mitigate the problem -- the Grid job automatically finds a working CE using the information system of the Grid. Naturally, if a Grid job is already running on the CE then a hardware error can be fatal for the job.
    \item {\bf Software management on the Grid.} There is no good and general solution for that. There is a possibility to send an executable file together with the job to the Grid, but it is the reasonable solution only if the executable file is small and does not have many external dependencies. Many international Grid projects are working in that field and hopefully some general solutions are coming soon. Additionally there are some problems connected to the commercial licensing politics.
    \item {\bf Missing accounting and banking system of computing time.} The problem is very urgent and needs a quick solution. For example, so called SGAS \cite{SGAS} that is developed by the SweGrid project \cite{SweGrid} can be one solution.
    \item {\bf Missing general job management tool for the users.} It is very complicated to manage a lot of running Grid jobs at the same time: resubmitting filed or uncorrect jobs, collecting data, etc. A solution can be the universal job manager software produced by Jensen {\it et al} \cite{JobManager}.
    \item {\bf Missing interface for the electronic ID-card (SmartCard).} The interface for the SmartCard/ID-card is under development by the authors.
    \item {\bf The organizational structure of the EG.} The Grid is a distributed system and it does not need to be highly centralized. However, at least some technical and political agreements are needed: the trust of the CA, exchange mechanisms and rates of the computational time, etc.
\end{itemize}

As Grid technology is a new and innovative topic in computational science and engineering, more courses and schooling are needed at the universities and other institutions. Additionally, the Grid is an international system and some interstate agreements are needed.

\section{Acknowledgements}
The authors would like to thank the team of the NorduGrid project for many very helpful suggestions. The authors like to thank Arto Teräs and Juha Lento at CSC in Helsinki for helping to set up the first components of the EG and giving a nice seminar at NICPB in Tallinn; Kristjan Kannike, Endel Lippmaa and Aleksander Trummal for very useful comments on the paper. The authors give a very warm thanks to Martti Raidal for the helpful suggestions and discussions in high energy and particle physics and Eero Vainikko for the valuable suggestions and discussions in scientific computations.

This work has been partly supported by the following organizations and institutions:
\begin{itemize}
    \item the Estonian Information Technology Foundation for the support for the completion the PC clusters at the University of Tartu,
    \item the Estonian Science Foundation for the grants no. 5135, 5935 and 5316,
    \item the EU 6th Framework Programme for the grant EC MC MERG-CT-2003-503626,
    \item the Ministry of Education and Research of Estonia for the support for the technical meetings of the EG,
    \item the Research Council of Norway (Nordplus Neighbour Programme) for the support for the NorduGrid technical meetings,
    \item the Swedish Institute (Visby Programme) for the support for Estonian students to study Grid technology at the Lund University.
\end{itemize}

\newpage

\begin{figure}
   \centering
   \includegraphics{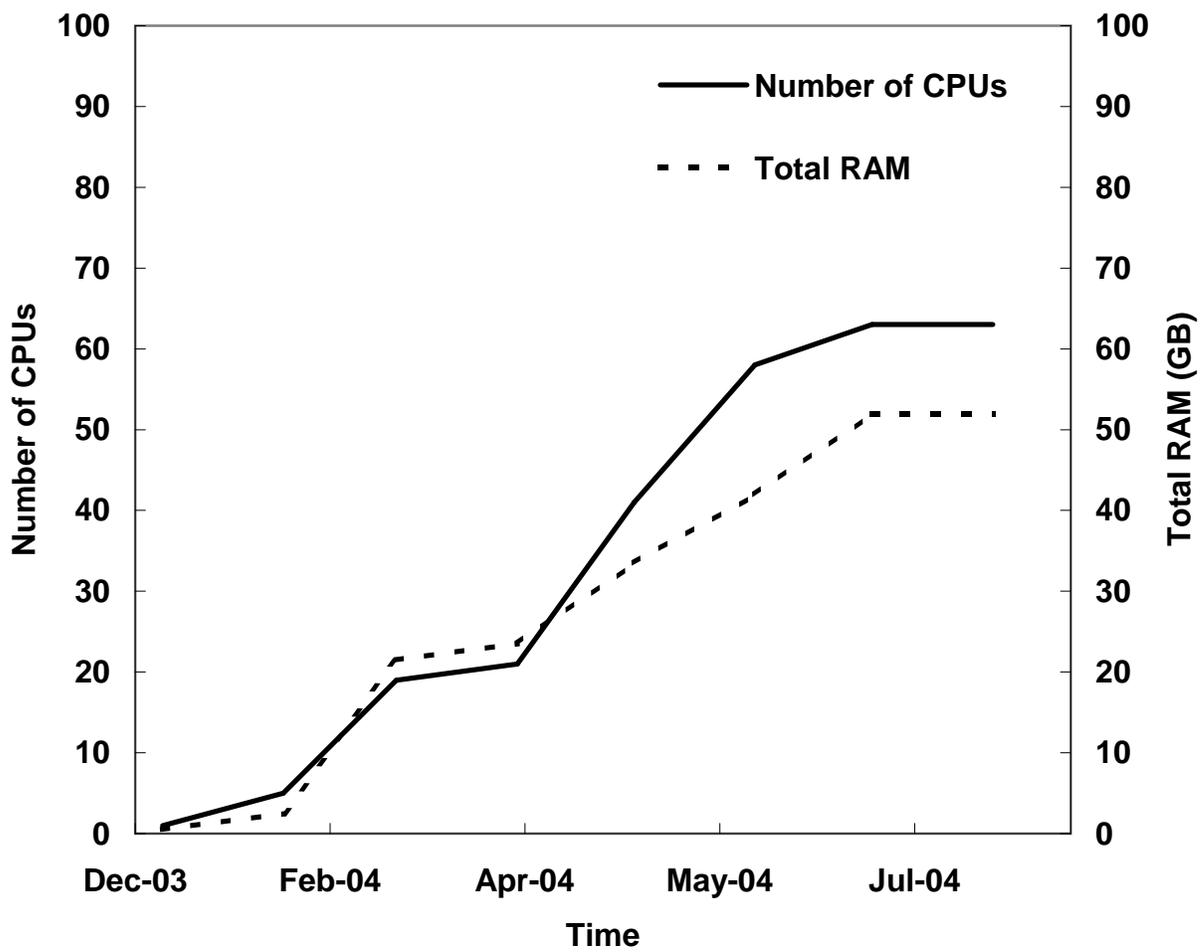}
   \caption{The number CPUs and RAM in the EG. Some new clusters are coming in the autumn 2004.}
   \label{fig:EGdevelopment}
\end{figure}

\begin{figure}
   \centering
   \includegraphics{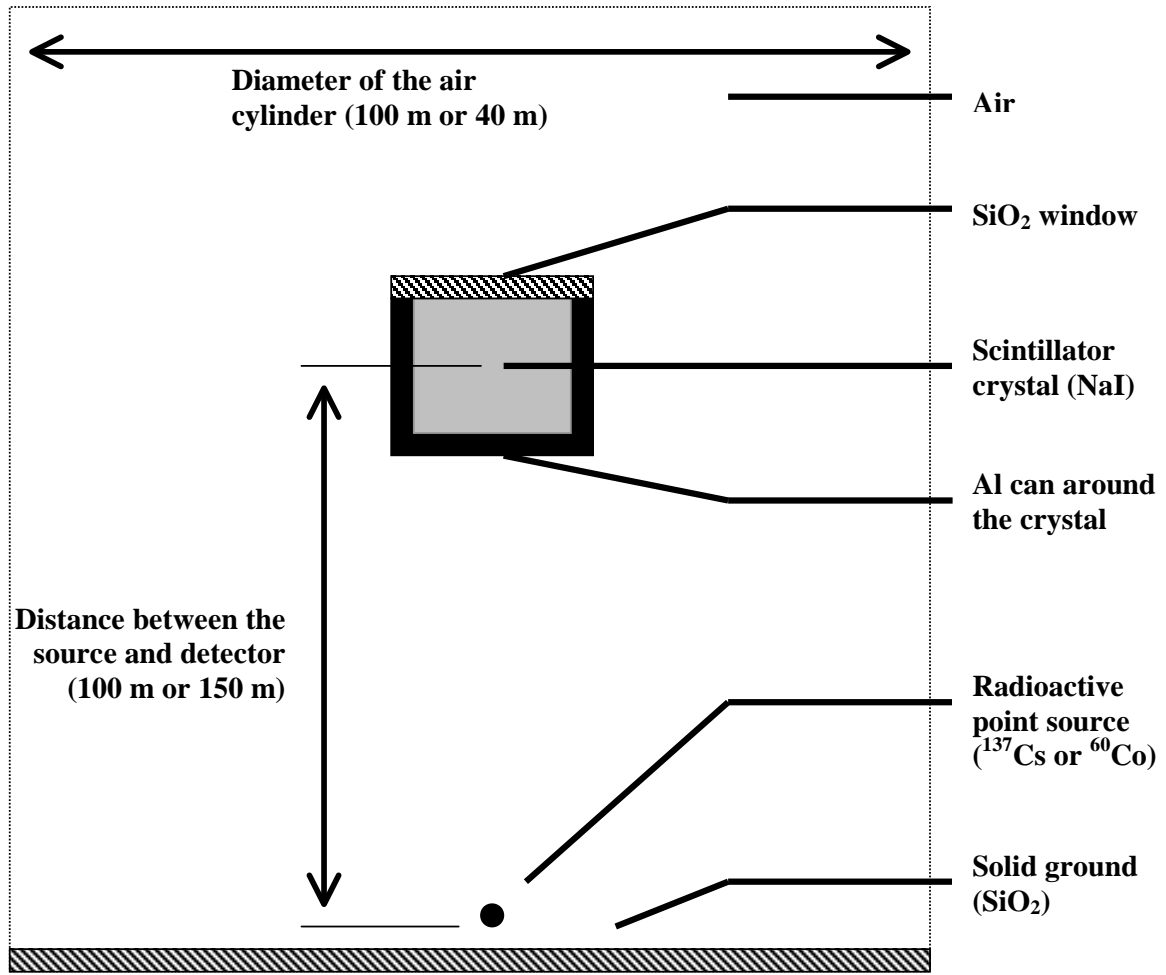}
   \caption{The schematic cross section of the simulated system. The detector crystal is cylinder, the diameter is about 15~cm and the height is about 5 cm.}
   \label{fig:stuk-system}
\end{figure}

\begin{figure}
   \centering
   \includegraphics{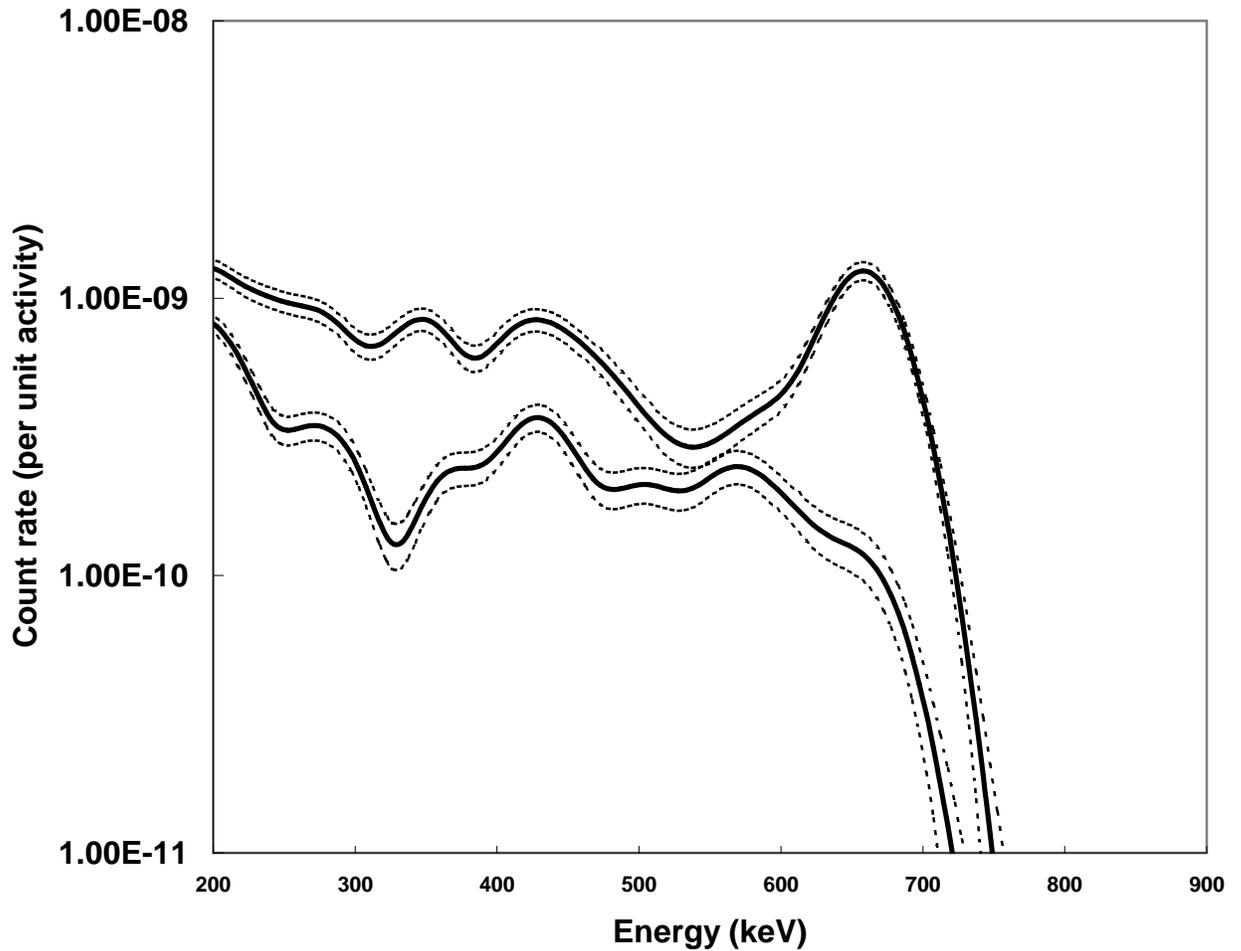}
   \caption{Two sets of spectra of $^{137}$Cs for the heights of 100 and 150 m in logarithmic scale. The upper solid curve presents the height of 150~m and the lower one presents that of 100~m. We can clearly see the energy peak on the upper curve at the energy of 662~keV. The Compton continuum region has the line structure. The reason of the line structure needs additional studies.}
   \label{fig:stuk-Cs}
\end{figure}

\begin{figure}
   \centering
   \includegraphics{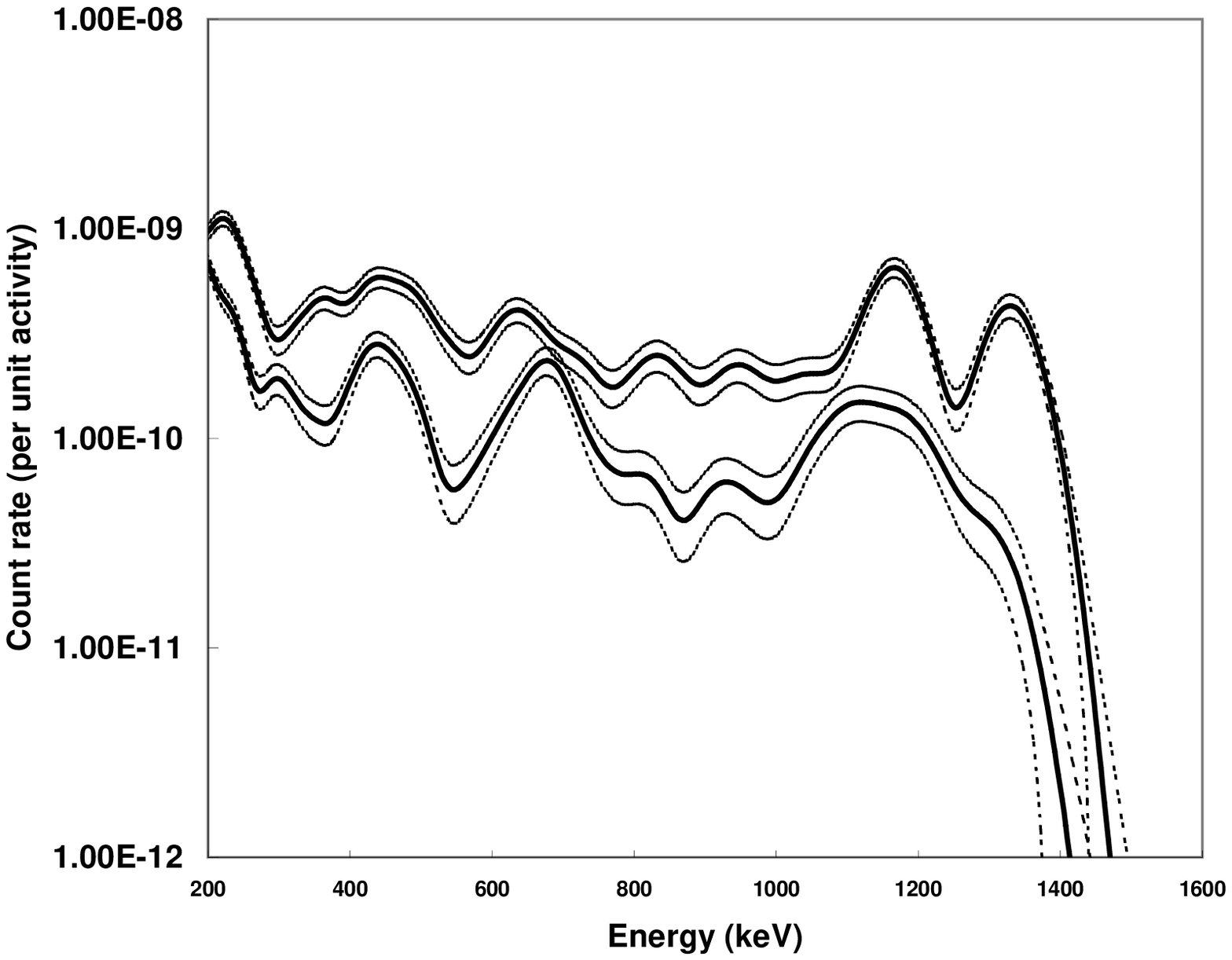}
   \caption{Two sets of spectra of $^{60}$Co for the heights of 100 and 150 m in logarithmic scale. The upper solid curve presents the height of 150~m and the lower one presents that of 100~m. We can clearly see the energy peaks on the upper curve at the energy of 1173.2~keV and 1332.5~keV. The Compton continuum region has the similar structure as the calculated spectra of $^{137}$Cs.}
   \label{fig:stuk-Co}
\end{figure}


\begin{thebibliography}{99}
\bibitem{InfFut:1994}
{\it Realizing the Information Future: The Internet and Beyond}.
National Academy Press, 1994
[http://www.nap.edu/readingroom/books/rtif/].
\bibitem{Foster:2001}
Foster~I., Kesselman~C. and Tuecke~S., The Anatomy of the Grid -
Enabling Scalable Virtual Organizations, {\it Int. J.
Supercomputer Applications}, 2001, {\bf 15}, 3-23
[arXiv:cs/0103025].
\bibitem{Foster:2003}
Foster~I. and Kesselman~C. {\it The Grid 2: Blueprint for a New
Computing Infrastructure}. Morgan Kaufmann, 2003.
\bibitem{Berman:2003}
Berman~F., Fox~G. and Hey~A.J.G. {\it Grid Computing: Making The Global
Infrastructure a Reality}. John Wiley \& Sons, 2003.
\bibitem{Asadzadeh:2004}
Asadzadeh~P., Buyya~R., Kei~C.L., Nayar~D. and Venugopal~S., Global Grids
and Software Toolkits: A Study of Four Grid Middleware
Technologies, {\it High Performance Computing: Paradigm and
Infrastructure, edited by Laurence Yang and Minyi Guo}. Wiley
Press, 2004 (in print) [arXiv:cs/0407001].
\bibitem{CERNexp}
European Organization for Nuclear Research [http://www.cern.ch/].
\bibitem{PlancMis}
The Planck Mission\\[0mm]
[http://www.rssd.esa.int/index.php?project=PLANCK].
\bibitem{HumanGenomeProject}
Human Genome Project\\[0mm]
[http://www.ornl.gov/sci/techresources/Human-Genome/home.shtml].
\bibitem{CERNandGrid}
LHC Computing Grid Project [http://lcg.web.cern.ch/LCG/].
\bibitem{GGF}
Global Grid Forum [http://www.ggf.org/].
\bibitem{Kadastik:2004}
Kadastik~M. {\it LHC Physics and Grid computing for CMS}.
Bachelor Thesis, 2004 [http://www.nicpb.ee/$\sim$mario/bsc.pdf].
\bibitem{GlobusToolkit}
The Globus Alliance [http://www.globus.org/].
\bibitem{GridAndPKI}
Ellison~M.C., The nature of a useable PKI, {\it Comp. Netw.}, 1999, {\bf 31}, 823-830.
\bibitem{EstIDcard}
The Estonian ID Card and Digital Signature Concept, 2003\\[0mm]
[http://www.id.ee/file.php?id=122].
\bibitem{Martens}
Martens~T., private communication.
\bibitem{EGWeb}
Estonian Grid Project [http://grid.eenet.ee/].
\bibitem{EGTechSup}
Technical Coordination and Support Group of the Estonian Grid\\[0mm]
[http://grid.eenet.ee/main.php?act=eesti-grid\&sact=tehniline\_tugi].
\bibitem{NorduGrid}
Ellert~M. {\it et al.}, The NorduGrid project: Using Globus toolkit for building Grid infrastructure, {\it Nucl. Instr. and Methods A}, 2003, {\bf 502} 407-410 [http://www.nordugrid.org/].
\bibitem{EuropeanDataGrid}
Kunszt~P., European DataGrid project: Status and plans, {\it Nucl. Instrum. Meth. A}, 2003, {\bf 502}, 376-382 [http://eu-datagrid.web.cern.ch/eu-datagrid/].
\bibitem{UNICORE}
UNICORE (UNiform Interface to COmputing REsources) Project [http://www.unicore.org/].
\bibitem{X.509}
Internet X.509 Public Key Infrastructure: Certificate and CRL Profile [http://www.ietf.org/rfc/rfc2459.txt].
\bibitem{EGCA}
Certification Authority of the Estonian Grid [http://grid.eenet.ee/].
\bibitem{EUGridPMA}
The European Policy Management Authority for Grid Authentication in e-Science [http://www.eugridpma.org/].
\bibitem{InfoSystemOfNorduGrid}
The NorduGrid Information System\\[0mm]
[http://www.nordugrid.org/documents/ng-infosys.pdf].
\bibitem{NGMonitor}
NorduGrid Monitor [http://www.nordugrid.org/monitor/].
\bibitem{EGMonitor}
Estonian Grid Monitor [http://giis.eenet.ee/monitor/].
\bibitem{CMS}
Compact Muon Solenoid\\[0mm]
[http://greybook.cern.ch/programmes/experiments/CMS.html].
\bibitem{CMSOO}
CMS Object-Oriented Projects\\[0mm]
[http://cmsdoc.cern.ch/cmsoo/cmsoo.html].
\bibitem{PYTHIA}
Sj\"ostrand~T., Ed\`en~P., Friberg~C., L\"onnblad~L., Miu~G., Mrenna~S. and Norrbin~E., High-Energy-Physics Event Generation with PYTHIA 6.1, {\it Computer Phys. Commun.} 2001, {\bf 135}, 238-256 [arXiv:hep-ph/0010017, http://www.thep.lu.se/$\sim$torbjorn/Pythia.html].
\bibitem{DOUG}
DOUG (Domain decomposition On Unstructured Grids)\\[0mm]
[http://www.maths.bath.ac.uk/$\sim$parsoft/doug/].
\bibitem{DOUGWeb}
The Web-Interface for the DOUG\\[0mm] [http://www.ce.ut.ee/$\sim$konstan/doug-www/].
\bibitem{DOUGamst}
Tehver~M., Vainikko~E., Skaburskas~K. and Vedru~J. Remote Access and Scheduling for Parallel Applications on Distributed Systems. -- Computational Science -- Lecture Notes in Computer Science, Sloot~P~M~A, Kenneth Tan~C~J, Dongarra~J~J, Hoekstra~A~G eds., Proc. ICCS 2002 Int. Conf., Springer, 2002, 633-642.
\bibitem{DOUGGrid}
The Grid-enabled Web-interface for the DOUG\\[0mm] [http://doug1.ce.ut.ee/doug/index.php]
\bibitem{MyProxy}
MyProxy - Online Credential Repository\\[0mm] [http://grid.ncsa.uiuc.edu/myproxy/]
\bibitem{EGMyProxies}
EG MyProxy servers: testsite.eenet.ee:7512, doug1.ce.ut.ee:7512.
\bibitem{Geant4}
Agostinelli~S {\it et al.}, GEANT4: A simulation toolkit, {\it Nucl. Instrum. Meth. A}, 2003, {\bf 506}, 250-303.
\bibitem{OpenMolGRID}
Open Computing Grid for Molecular Science and Engineering (OpenMolGRID) [http://www.openmolgrid.org].
\bibitem{RD44}
Geant4 Research and Development Project\\[0mm]
[http://pcitapiww.cern.ch/asd/geant/rd44.html].
\bibitem{STUK:2003}
Hektor~A. {\it et al.}, Monte Carlo simulation for a scintillation detector to be used in a prototype radiation surveillance unit of Ranger, {\it report for the Finnish Scientific Advisory Board for Defence (MATINE), available only in a special agreement with MATINE}.
\bibitem{STUKHektor:2004}
Hektor~A., Kurvinen~K., P\"oll\"anen~R. and Smolander~P., Geant4 simulations for a scintillation detector to be used in a prototype radiation surveillance unit, {\it under preparation}.
\bibitem{CLHEP}
Class Library for High Energy Physics\\[0mm]
[http://wwwasd.web.cern.ch/wwwasd/lhc++/clhep/].
\bibitem{HektorAarhus}
Hektor~A., Kolbe~E., Langanke~K., and Toivanen~J., Neutrino-induced reaction rates for r-process nuclei, {\it Phys. Rev. C}, 2000, {\bf 61}, 055803-055813.
\bibitem{HektorUppsala}
Hektor~A., Klintenberg~M.C., Aabloo~A. and Thomas~J.O., Molecular dynamics simulation of the effect of a side chain on the dynamics of the amorphous LiPF6–PEO system, {\it J. Matr. Chem.}, 2003, {\bf 13}, 214-218. 
\bibitem{SweGrid}
SweGrid [http://www.swegrid.se/].
\bibitem{SGAS}
SweGrid Accounting System Project\\[0mm]
[http://www.pdc.kth.se/grid/sgas/].
\bibitem{JobManager}
Job Manager for the NorduGrid ARC\\[0mm]
[http://www.cs.aau.dk/$\sim$htj/nordugrid/master\_thesis.ps].
\end{thebibliography}
\end{document}